\documentclass[aps,english,showpacs,twocolumn]{revtex4-1}
\usepackage{amssymb}
\usepackage{amsmath}
\usepackage{graphicx}
\usepackage{epsfig}
\usepackage{color}

\begin{document}

\title{Parity-time symmetric coupled asymmetric dimers}
\author{L. Jin}
\email{jinliang@nankai.edu.cn}
\affiliation{School of Physics, Nankai University, Tianjin 300071, China}

\begin{abstract}
{We investigate a parity-time ($\mathcal{PT}$) symmetric system that
consists of two symmetrically coupled asymmetric dimers. The enclosed
magnetic flux controls the $\mathcal{PT}$ phase transition. The system can
reenter the exact $\mathcal{PT}$-symmetric phase from a broken $\mathcal{PT}$%
-symmetric phase with large non-Hermiticity. Two-state coalescence may have
one or two defective eigenstates. The topology of exceptional points is
reflected by the magnetic flux independent phase rigidity scaling exponents.
The topology changes when exceptional points coincide. The geometric phases
accumulate when encircling the exceptional points and vary as the magnetic
flux. The magnetic flux is favorable for the realization of high-order
exceptional points. A triple point of different quantum phases has an order
of four. The perturbation around the four-state coalescence leads to a
fourth-root mode frequency splitting; the sensing sensitivity is
significantly enhanced.}
\end{abstract}

\maketitle


\section{Introduction}

Parity-time ($\mathcal{PT}$) symmetric non-Hermitian Hamiltonians can
possess real spectra; this discovery stimulated a burst of research interest
in the extension of quantum mechanics~\cite%
{Bender,Dorey,Bender02,Ali,Jones,Znojil,LJin09,LJin10,Witthaut,YNJ}. $%
\mathcal{PT}$ symmetry is the origin of many intriguing features in
non-Hermitian systems. Various optical systems are fruitful platforms for
the investigation of $\mathcal{PT}$ symmetry~\cite{AR,OL,Musslimani,Klaiman}%
. The phenomena of open quantum systems can be described by $\mathcal{PT}$%
-symmetric non-Hermitian Hamiltonians after an average overall decay has
been removed~\cite{Rotter09,Rotter15}. In a passively coupled waveguide,
where losses were asymmetric, $\mathcal{PT}$ symmetry breaking was
demonstrated~\cite{AGuo}; the $\mathcal{PT}$-symmetric coupled waveguide was
realized through introducing an active gain; the light intensity oscillated
in the exact $\mathcal{PT}$-symmetric phase and exponentially increased in
the broken $\mathcal{PT}$-symmetric phase~\cite{CERuter}. $\mathcal{PT}$
symmetry was also proposed and experimentally realized in coupled
microcavities~\cite{Jing,PengNP,PengScience,FengScience,Chang}. In a linear region of an
active cavity, doped ions under pumping generated a self-adapted net gain
that balanced the loss in the corresponding dissipative cavity. The light
transport is reciprocal and nonreciprocal in the exact and broken $\mathcal{%
PT}$-symmetric phases, respectively~\cite{PengNP}. When the gain saturation
induces large nonlinearity dominates, the coupled microcavities are
nonreciprocal and can be applied as an optical isolator~\cite{Chang}.

Recently, research has addressed the exceptional points (EPs). The EP in a $%
\mathcal{PT}$-symmetric system is the $\mathcal{PT}$ phase transition point.
The $\mathcal{PT}$-symmetric systems with periodical potentials have also
been investigated. Unidirectional reflection suppression was demonstrated at
the $\mathcal{PT}$ phase transition point~\cite{Feng}. The $\mathcal{PT}$
phase transition threshold exists in a $\mathcal{PT}$-symmetric sinusoidal
potential, but any higher degree of non-Hermiticity leads to $\mathcal{PT}$
symmetry breaking~\cite{GraefePRA}. The competition between two lattice
potentials can induce $\mathcal{PT}$ symmetry breaking and restoration as
non-Hermiticity increasing~\cite{YNJAA}. EPs enhance the optical sensing~%
\cite{Wiersig,ChenSensing}. State flip or mode switch occurs when encircling the EPs sufficiently slowly for integer circles. The geometric phases are accumulated in the encircling process and the intriguing topologies of EPs are reflected~\cite{Dembowski,Uzdin,WDHeissEP,Menke}. In a two-level system, when encircling a two-state coalescence exceptional point (EP2), the two eigen energy levels switch to each other after encircling the EP2 for one circle; and the geometric phase accumulated for each eigenstate is $\pi$ after encircling the EP2 for two circles; to make each eigenstate back to itself, encircling EP2 for four circles is needed~\cite{Doppler,Xu}. High-order EPs have highly complicated topological structures and physical
implications~\cite{CTChenPRX}, the sensitivity is enhanced due to the cubic-root frequency response near a three-state coalescence exceptional point~\cite{HodaeiSensing}.

Photons, as neutral particles, do not directly interact with magnetic
fields. Recent studies have implemented artificial magnetic fields for
photons through dynamic modulation of material permittivity, optomechanical
coupling, and photon-phonon interactions~\cite{Fang,Fang2,Tzuang,Hafezi,Li}.
The artificial magnetic field provides a new degree of freedom, which is
favorable for optical control and manipulation. In this work, we study $%
\mathcal{PT}$-symmetric coupled asymmetric dimers, where effective magnetic
flux is induced by the nonreciprocal coupling of the dimers. We demonstrate
that the energy level structure, the quantum phases, and the topology of EPs
are affected by the magnetic flux, the coupling, and the degree of
non-Hermiticity. We find that the system can reenter an exact $\mathcal{PT}$%
-symmetric phase from a broken $\mathcal{PT}$-symmetric phase at large
non-Hermiticity. The gap between two central energy levels vanishes at
uniform coupling in the absence of magnetic flux; the chirality of the EPs
depends on the competition between the two coupling strengths. EP2 with two
defective eigenstates exists, where a pair of two-state coalescences appear.
Phase rigidity is a useful measure of state mixing; its scaling exponent
characterizes the topology of EPs, which is magnetic-flux-independent. The
topology of EPs changes when they coincide. Four-state coalescence (EP4)
appears when EP2 with one and two defective eigenstates coalesces, the
perturbation around EP4 leads to a fourth-root mode frequency splitting; in
that state, sensing sensitivity is substantially enhanced.

\section{$\mathcal{PT}$ symmetric asymmetric dimers}

\begin{figure}[tb]
\includegraphics[ bb=195 375 380 585, width=4 cm, clip]{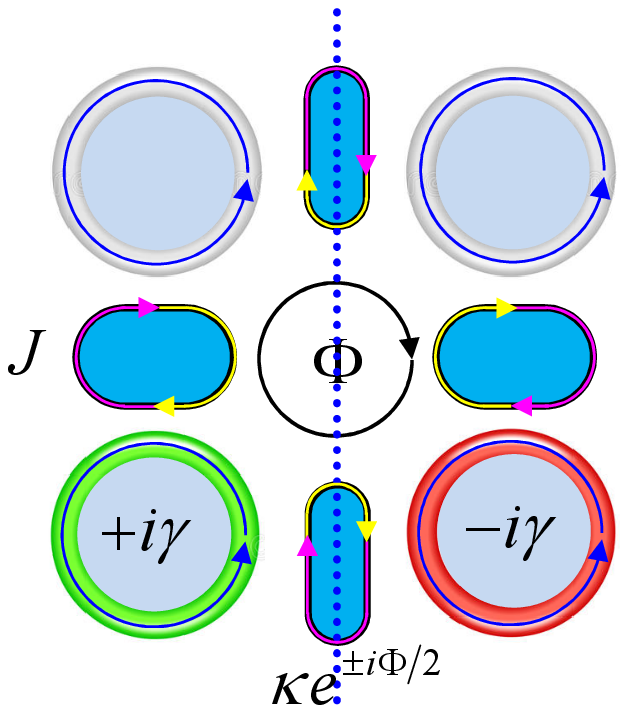}
\caption{Schematic of a four coupled
resonators consisting of two asymmetric dimers. One dimer has loss (red) and the other has gain (green). The coupling strengths are $\kappa$ and $J$. The dotted blue line is the $\mathcal{PT}$-symmetric axis. The primary (auxiliary) resonators are round (elliptical). The blue arrows indicate the counterclockwise mode supported by the primary resonators. The yellow (magenta) arrows illustrate the path lengths that counterclockwise mode photons traveling in counterclockwise (clockwise) direction between primary resonators. The length difference induces a directional phase factor in the coupling, which results in the nonreciprocity and induces the synthetic magnetic flux $\Phi$.
} \label{fig1}
\end{figure}

We consider two asymmetric dimers symmetrically coupled in a closed configuration, one dimer has gain (green) and the other has loss (green).
The system is $\mathcal{PT}$-symmetric with respect to the left-right
reflection, as schematically illustrated in Fig.~\ref{fig1}. The parity operation is
defined as the space reflection with respect to the $\mathcal{PT}$-symmetric
axis (blue dotted line in Fig.~\ref{fig1}).

We investigate the influence of
nonreciprocal couplings that induce effective magnetic flux on the spectrum,
quantum phases, and the topology of EPs. Notably, the magnetic
flux as an additional degree of freedom does not break the $\mathcal{PT}$
symmetry of the system. The magnetic flux provides an alternative way for
the control of $\mathcal{PT}$-symmetric phases and EPs, which may facilitate the application
of $\mathcal{PT}$ metamaterials.

In the coupled resonator system, all the primary resonators possess identical resonant frequency. They are evanescently
coupled through auxiliary resonators, the frequency of which is antiresonant
with the primary resonators~\cite{MHafezi}. The optical path length
difference introduced through the coupling process induces a nonreciprocal
coupling phase factor, which effectively realizes a synthetic magnetic
flux in the closed configuration formed by the four coupled primary resonators. The nonreciprocal extra phase factor in the coupling is $e^{\pm 2\pi \Delta x/\lambda }$ for the path length difference $2\Delta x$, where $\lambda $ is the
resonant wavelength~\cite{HafeziPRL}. The inter-dimer (horizontal) coupling strength is $%
\kappa $ with nonreciprocal phase factor $e^{\pm i\Phi /2}$, the intra-dimer
(vertical) coupling strength is $J$. The synthetic magnetic flux introduced in the coupled four resonator system equals to $\Phi=4\pi \Delta x/\lambda$.
The gain in the resonator is based on a linear model by assuming a certain gain rate $\gamma$~\cite{FengScience}. The gain in one dimer equals to the loss in the other dimer to form a $\mathcal{PT}$-symmetric system.

The system is described by a $4\times 4$
Hamiltonian,
\begin{equation}
H=\left(
\begin{array}{cccc}
-i\gamma & \kappa e^{i\Phi /2} & 0 & J \\
\kappa e^{-i\Phi /2} & +i\gamma & J & 0 \\
0 & J & 0 & \kappa e^{i\Phi /2} \\
J & 0 & \kappa e^{-i\Phi /2} & 0%
\end{array}%
\right) .  \label{H}
\end{equation}%
The eigen energy of Hamiltonian $H$ is $E_{\pm ,\pm }=\pm \sqrt{\Delta
_{1}\pm \sqrt{\Delta _{2}}}$ with $\Delta _{1}=\kappa ^{2}+J^{2}-\gamma
^{2}/2$ and $\Delta _{2}=4J^{2}\kappa ^{2}\cos ^{2}(\Phi /2)-J^{2}\gamma
^{2}+\gamma ^{4}/4$.

\section{$\mathcal{PT}$ symmetric phase}
\begin{figure}[tb]
\includegraphics[ bb=0 0 330 490, width=8.6 cm, clip]{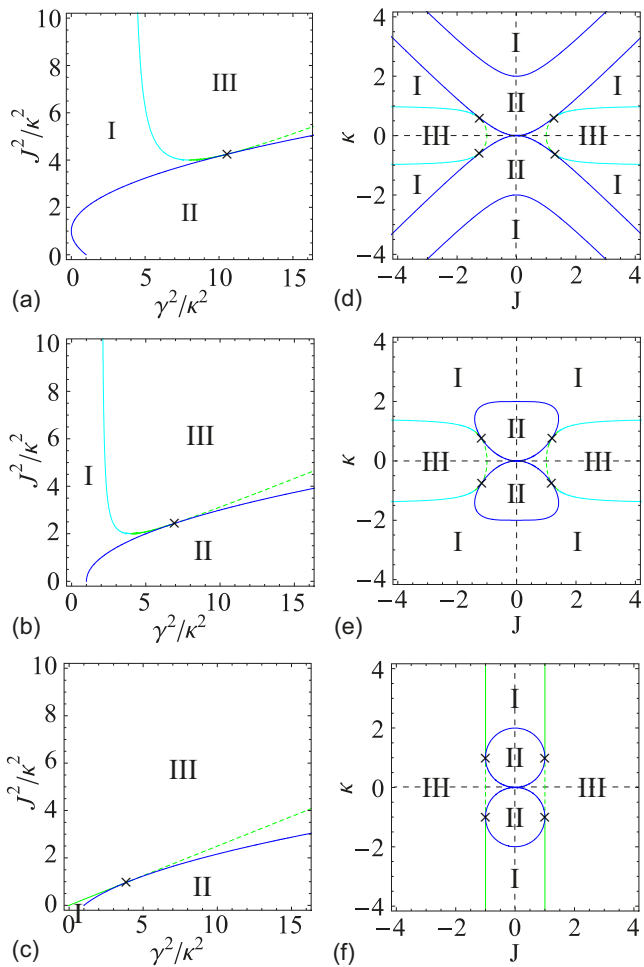}
\caption{(a,d) $\Phi=0$, (b,e) $\Phi=\pi/2$, and (c,f) $\Phi=\pi$. $\mathcal{PT}$ phase diagrams are shown in the $\gamma^2$-$J^2$ plane for (a-c).
I is the exact $\mathcal{PT}$-symmetric phase, II and III are the broken $\mathcal{PT}$-symmetric phases with one and two pairs of conjugate energy levels, respectively. The critical lines of $\gamma_{\rm{c,1}}$ and $\gamma_{\rm{c,2,\pm}}$ are shown in the $\kappa$-$J$ plane for (d-f); the couplings are in the unit of a decay rate, $\gamma/2=1$. The dashed green curve divides
the pure imaginary and complex energy levels in region III. The blue solid curves ($\gamma_{\rm{c,1}}$) represent the EP2 with one defective eigenstate; the cyan ($\gamma_{\rm{c,2,-}}$) and green curves ($\gamma_{\rm{c,2,+}}$), both the solid and dashed ones, indicate the EP2 with two defective eigenstates; black crosses mark the high-order EP4s where two types of EP2s coincide.}
\label{fig2}
\end{figure}
For $\Delta _{2}>0$, $\Delta _{1}>\sqrt{\Delta _{2}}$, the four energy
levels are all real (phase I). For $\Delta _{2}>0$ and $\Delta
_{1}^{2}<\Delta _{2}$, two energy levels $\pm \sqrt{\Delta _{1}+\sqrt{\Delta
_{2}}}$ are real, and the other two energy levels $\pm \sqrt{\Delta _{1}-%
\sqrt{\Delta _{2}}}$ form a conjugate pair (phase II). For $\Delta _{2}<0$
or $\Delta _{1}<0$ and $\Delta _{1}^{2}>\Delta _{2}>0$, four energy levels
form two conjugate pairs (phase III). In Fig.~\ref{fig2}, the phase diagram
is plotted for different magnetic fluxes. The blue and cyan
curves constitute a $\lambda $-curve; on the left side of the $\lambda $%
-curve is the exact $\mathcal{PT}$-symmetric phase with an entirely real
spectrum (phase I); on the right side of the $\lambda $-curve is the broken $%
\mathcal{PT}$-symmetric phase, which includes one pair (two pairs) of
conjugate energy levels in phase II (III). Phase II is a partially broken $%
\mathcal{PT}$ phase; phase III is a completely broken $\mathcal{PT}$ phase.
The solid curves are the boundary between different phases. The cyan and green curves represent $\Delta _{2}=0$, which corresponds to the EP2 with a pair of two-state coalescences; the system has two defective eigenstates in this situation. The blue curves represent $\Delta _{1}^{2}=\Delta _{2}$, which corresponds to the EP2 with a two-state coalescence; the system has one defective eigenstate in this situation. The
solid green curves in Fig.~\ref{fig2} are the boundary between region I and III, where $\mathcal{PT}$ symmetry breaking occurs. The dashed green curves in Fig.~\ref{fig2} are the boundary between energy levels being pure imaginary and being complex numbers. In the presence of magnetic flux, both critical coupling and degree of
non-Hermiticity decrease. Thus, the synthetic magnetic flux is favorable for
the realization of $\mathcal{PT}$ phase transition and high-order EPs.

To have real energy, $\Delta _{2}\geqslant 0$ is necessary. At $\Delta
_{2}=0 $, the system is at an EP2 with a pair of two-state coalescences,
this happens at%
\begin{equation}
\gamma _{\mathrm{c,2,\pm }}^{2}=2J^{2}\pm 2\sqrt{J^{4}-4J^{2}\kappa ^{2}\cos
^{2}(\Phi /2)},
\end{equation}%
where the upper two levels and the lower two levels coalesce. The system has
two defective eigenstates and its spectrum is a coalesced conjugate pair.
Notably, $\gamma _{\mathrm{c,2,-}}=\gamma _{\mathrm{c,2,+}}$ at
\begin{equation}
J_{\mathrm{c,2}}^{2}=4\kappa ^{2}\cos ^{2}(\Phi /2).  \label{Jc2}
\end{equation}%
When $J^{2}<J_{\mathrm{c,2}}^{2}$, the EP2s with a pair of two-state
coalescences disappear, and we have $\Delta _{2}>\left( J^{2}-\gamma
^{2}/2\right) ^{2}\geqslant 0$. When $J^{2}>J_{\mathrm{c,2}}^{2}$, then $%
\Delta _{2}>0$ requires $\gamma ^{2}<\gamma _{\mathrm{c,2,-}}^{2}$ or $%
\gamma ^{2}>\gamma _{\mathrm{c,2,+}}^{2}$. Notably, $J^{2}>J_{\mathrm{c,2}%
}^{2}$ is always satisfied for magnetic flux $\Phi =2n\pi +\pi $ ($n\in
\mathbb{Z}
$), where $\gamma _{\mathrm{c,2,-}}$ vanishes and the system is fragile to
non-vanishing gain and loss.

The couplings change the weight of the eigenstate probability distribution.
At any weak $J$ ($J^{2}<J_{\mathrm{c,2}}^{2}$), the four resonators are
considered as a $\mathcal{PT}$-symmetric dimer (lower two resonators) weakly
coupled to a Hermitian dimer (upper two resonators). The upper and lower
levels are always real. Any non-Hermiticity diminishes the mode splitting of
the $\mathcal{PT}$-symmetric dimer; an EP2 $\gamma _{\mathrm{c,1}}$ with two
central energy levels coalesced at energy $0$ appears at large degree of
non-Hermiticity. At strong $J$ ($J^{2}>J_{\mathrm{c,2}}^{2}$), the system is
considered as two asymmetric dimers coupled through weak nonreciprocal
coupling $\kappa e^{\pm i\Phi /2}$; the left and right dimers are both
non-Hermitian, describing by an identical $\mathcal{PT}$-symmetric dimer
after removing the overall decay factors $+i\gamma /2$ and $-i\gamma /2$.

When $\Delta _{1}^{2}=\Delta _{2}\neq 0$, the central two energy levels
coalesce at $E=0$ and the system is at an EP2 of a two-state coalescence
with one defective eigenstate. The critical gain and loss must satisfy%
\begin{equation}
\gamma _{\mathrm{c,1}}^{2}=\left( J^{2}-\kappa ^{2}\right) ^{2}/\kappa
^{2}+4J^{2}\sin ^{2}(\Phi /2).
\end{equation}%
At $\Phi =2n\pi $ ($n\in
\mathbb{Z}
$), the eigenstate for $E=0$ is $\psi _{\sigma }=(-i\sigma \kappa /J,-\kappa
/J,i\sigma ,1)^{T}$ with $\sigma =\left\vert J^{2}-\kappa ^{2}\right\vert
/(J^{2}-\kappa ^{2})$. For the two-state coalescence EP2 with one defective
eigenstate, encircling the EP2 in the parameter space with one circle
induces the eigenstates to switch~\cite{Lee}; a geometric phase of $\pm \pi $
is accumulated when encircling the EP2 with two circles that have different
chiralities~\cite{WDHeissEP}. At $\Phi =2\pi +2n\pi $ ($n\in
\mathbb{Z}
$), the eigenstate for $E=0$\ is $\psi _{\sigma }=(-i\sigma \kappa /J,\kappa
/J,-i\sigma ,1)^{T}$. The chirality of EP2 is left when $\sigma $ is $+1$
for $J^{2}>\kappa ^{2}$; and the chirality of EP2 is right when $\sigma $ is
$-1$ for $J^{2}<\kappa ^{2}$. The chirality of the EPs depends on the
competition between the two coupling strengths. The magnetic flux does not
change the chirality of EP2s.

The gap between the central two levels closes at $\gamma =0$ for $\left(
\kappa -J\right) ^{2}+4J\kappa \sin ^{2}(\Phi /2)=0$. That is only true when
$J=\kappa $, and $\Phi =2n\pi $ ($n\in
\mathbb{Z}
$). For trivial magnetic flux $\Phi =2n\pi $ ($n\in
\mathbb{Z}
$), the critical lines of $\gamma _{\mathrm{c,1}}$ are open curves in the $J$%
-$\kappa $ plane [blue curves in Fig.~\ref{fig2}(d)], which
are divided into the left chiral EP and the right chiral EP. For nontrivial
magnetic flux $\Phi \neq 2n\pi $ ($n\in
\mathbb{Z}
$), the critical lines of $\gamma _{\mathrm{c,1}}$ are closed circles in the
$J$-$\kappa $ plane [blue circles in Fig.~\ref{fig2}(e,f)].

There exist two types of $\gamma _{\mathrm{c,1}}$: Case I, $\Delta _{1}>0$, $%
\Delta _{1}-\sqrt{\Delta _{2}}=0$. The EP2 satisfies $\gamma _{\mathrm{c,1}%
}^{2}<2\kappa ^{2}+2J^{2}$, which requires $J^{2}<J_{\mathrm{c,1}}^{2}$,
\begin{equation}
J_{\mathrm{c,1}}^{2}=2\kappa ^{2}\cos ^{2}(\Phi /2)+\kappa ^{2}\sqrt{4\cos
^{4}(\Phi /2)+1}.  \label{Jc1}
\end{equation}%
Case II, $\Delta _{1}<0$, $\Delta _{1}+\sqrt{\Delta _{2}}=0$. The EP2
satisfies $\gamma _{\mathrm{c,1}}^{2}>2\kappa ^{2}+2J^{2}$, which requires $%
J^{2}>J_{\mathrm{c,1}}^{2}$. In case I, when the system parameter crosses $%
\gamma _{\mathrm{c,1}}$, the system experiences a $\mathcal{PT}$ phase
transition; while in case II, both two sides of $\gamma _{\mathrm{c,1}}$ are
$\mathcal{PT}$ symmetry broken, but with one and two pairs of complex energy
levels, respectively.
\begin{figure*}[t]
\includegraphics[ bb=0 0 550 180, width=18 cm, clip]{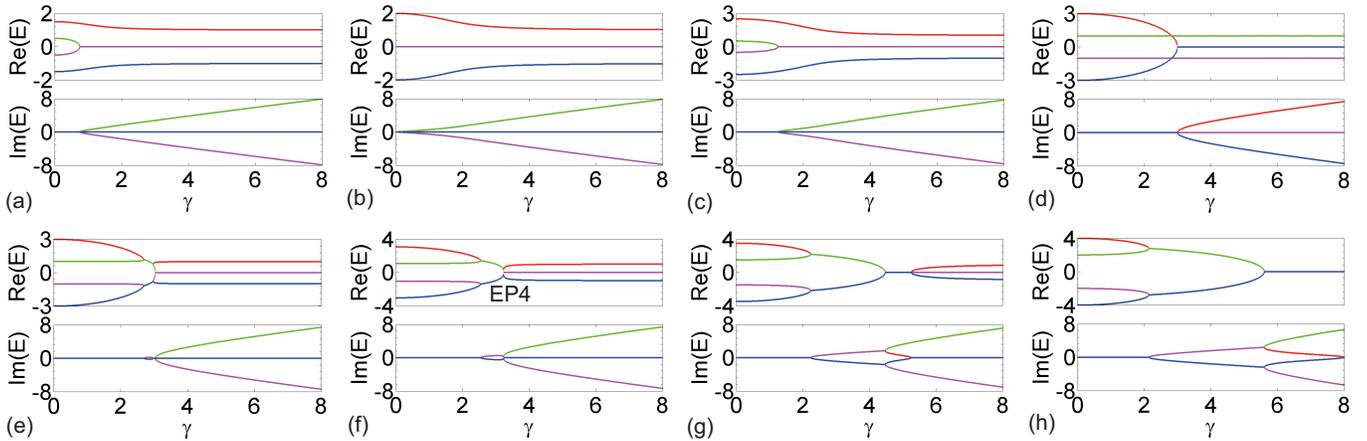}
\caption{Energy levels as functions of gain and loss $\protect\gamma $. (a) $%
J=0.5$, (b) $J=1.0$, (c) $J=1.5$, (d) $J=J_{\mathrm{c,2}}=2.0$, (e) $J=2.01$%
, (f) $J=J_{\mathrm{c,1}}=\protect\sqrt{2+\protect\sqrt{5}}\approx 2.058$,
(g) $J=2.5$, (h) $J=3.0$. Other parameters are $\protect\kappa =1$, $\Phi =0$%
.}
\label{fig3}
\end{figure*}

The system has one (two) defective eigenstate (eigenstates) at $\gamma _{%
\mathrm{c,1}}$ ($\gamma _{\mathrm{c,2,\pm }}$). The critical $\gamma _{%
\mathrm{c,2,-}}$ ($\gamma _{\mathrm{c,2,+}}$) is illustrated in cyan (green) in Fig.~\ref{fig2}(d-f). The curves $\gamma _{%
\mathrm{c,2,-}}$ and $\gamma _{\mathrm{c,2,+}}$ coincide at $\gamma _{%
\mathrm{c,2,-}}^{2}=\gamma _{\mathrm{c,2,+}}^{2}=2J^{2}$ when $J^{2}=J_{%
\mathrm{c,2}}^{2}$ [Eq. (\ref{Jc2})]; that is, when a pair of two-state
coalescences appear, the energy is $E_{\pm ,\pm }=\pm \kappa $. Note that
for $\gamma _{\mathrm{c,1}}^{2}\geqslant \gamma _{\mathrm{c,2,+}%
}^{2}\geqslant \gamma _{\mathrm{c,2,-}}^{2}$, the curves $\gamma _{\mathrm{%
c,2,+}}^{2}$ and $\gamma _{\mathrm{c,1}}^{2}$ coincide at $J^{2}=J_{\mathrm{%
c,1}}^{2}$ [Eq. (\ref{Jc1})], where four eigenstates coalesce at zero
energy; the system is at an EP4 with three defective eigenstates. The EP4 is
the triple point of three different phases: I, II, and III. The critical
gain and loss is $\gamma _{\mathrm{EP4}}^{2}=2\kappa ^{2}[1+2\cos ^{2}\left(
\Phi /2\right) +\sqrt{4\cos ^{4}\left( \Phi /2\right) +1}]$, which is $%
\gamma _{\mathrm{EP4}}^{2}=4\kappa ^{2}$ for $\kappa =J=1$ as depicted in
Fig.~\ref{fig2}(f) when $\Phi =\pi $. The only eigenstate at the EP4 is $%
(1,1,i,-i)^{T}$.

In a Hermitian system, perturbation leads to energy splitting of degenerate
states; the resulting energy splitting is proportional to the perturbation $%
\epsilon $\ around Hermitian degeneracies or diabolic points. Operating
around EPs or the non-Hermitian degeneracies, the energy splitting is more
sensitive to the perturbation~\cite{Wiersig}; in particular, the sensitivity
is significantly enhanced for tiny perturbations. The detection sensitivity
enhancement at two-order and three-order EPs have been demonstrated in
optical systems; the mode frequency splitting induced by the perturbation
scales as $\epsilon ^{1/N}$\ for an $N$-state coalescence in a non-Hermitian
system~\cite{ChenSensing,HodaeiSensing}. The sensitivity increases at
high-order EPs. In the asymmetric coupled dimers, EP4 exhibits a mode
frequency splitting $E\approx \pm \sqrt{\pm 2e^{i3\pi /4}}\epsilon ^{1/4}$
for perturbation $\epsilon $ on the gain and loss resonators ($i\gamma
\rightarrow i\gamma +\epsilon $). The fourth-root $E\propto \epsilon ^{1/4}$
results in a significant sensitivity enhancement.

\section{Energy level structure}

The spectrum is symmetric about zero energy; the energy levels exhibit rich
structures that are related to the EPs. In Fig.~\ref{fig3}, the energy
levels are depicted for different values of coupling $J$ at trivial magnetic
flux $\Phi =0$ as functions of gain and loss $\gamma $. In the plots from
Fig.~\ref{fig3}(a) to Fig.~\ref{fig3}(h), the coupling $J$ increases. The
coupling $J\leqslant J_{\mathrm{c,2}}=2\kappa $ is depicted in the upper
panel of Fig.~\ref{fig3}(a-d), where the system has two real energies. The
mode frequency (real part of the energy levels) decreases as gain and loss $%
\gamma $ increases; the EP $\gamma _{\mathrm{c,1}}$ appears as the central
two levels coalesce at $E_{\pm ,-}=0$. $\gamma _{\mathrm{c,1}}$ in Fig.~\ref%
{fig3}(a) has right chirality. As the coupling increases to $J=\kappa $, the
gap between the central two levels closes (Fig.~\ref{fig3}b), thereafter, $%
\gamma _{\mathrm{c,1}}$ in Fig.~\ref{fig3}(c) has left chirality. At $J=J_{%
\mathrm{c,2}}=2\kappa $, two $\gamma $-independent real energy levels $\pm
\kappa $ appear [Fig.~\ref{fig3}(d)].

For $J^{2}>J_{\mathrm{c,2}}^{2}$, the critical gain and loss satisfy $\gamma
_{\mathrm{c,2,-}}<\gamma _{\mathrm{c,2,+}}<\gamma _{\mathrm{c,1}}$. The real
parts of the uppermost and lowest levels decrease, the central two levels
increase with $\gamma $ until the upper and lower two energy levels coalesce
at EP2 $\gamma _{\mathrm{c,2,-}}$, respectively. When $J_{\mathrm{c,2}%
}^{2}<J^{2}<J_{\mathrm{c,1}}^{2}$, the upper and lower two energy levels
coalesce at $\gamma _{\mathrm{c,2,-}}$ where $\mathcal{PT}$ transition
occurs. The system changes from the $\mathcal{PT}$-symmetric phase into the
broken $\mathcal{PT}$-symmetric phase; the eigen energies become two complex
conjugate pairs. As $\gamma $ further increases, two pairs of conjugate
energy levels bifurcate at $\gamma _{\mathrm{c,2,+}}$ and the system enters
the $\mathcal{PT}$-symmetric phase once more until the central two levels
coalesce at $\gamma _{\mathrm{c,1}}$ [Fig.~\ref{fig3}(e)]; the system
reenters the exact $\mathcal{PT}$-symmetry region at larger degrees of
non-Hermiticity. Therefore, $\gamma <\gamma _{\mathrm{c,2,-}}$ and $\gamma _{%
\mathrm{c,2,+}}<\gamma <\gamma _{\mathrm{c,1}}$ are in the exact $\mathcal{PT%
}$-symmetric phase; $\gamma _{\mathrm{c,2,-}}<\gamma <\gamma _{\mathrm{c,2,+}%
}$ and $\gamma >\gamma _{\mathrm{c,1}}$ are in the broken $\mathcal{PT}$%
-symmetric phase. At $J^{2}=J_{\mathrm{c,1}}^{2}$, two types of EP2s $\gamma
_{\mathrm{c,2,+}}$ and $\gamma _{\mathrm{c,1}}$ coincide and all four levels
coalesce at $E_{\pm ,\pm }=0$ [Fig.~\ref{fig3}(f)].

\begin{figure}[t]
\includegraphics[ bb=0 0 545 360, width=8.6 cm, clip]{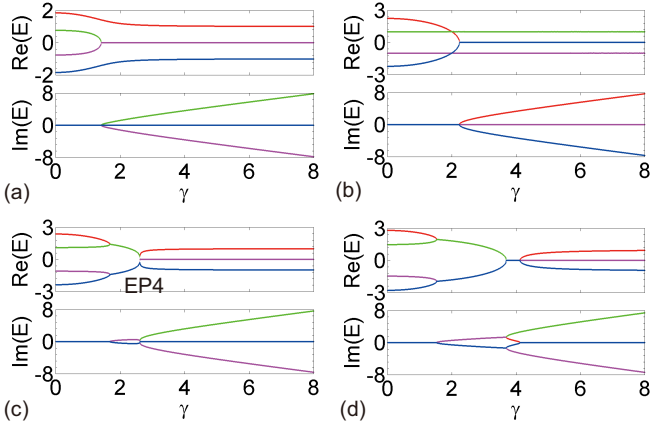}
\caption{Energy levels as functions of gain and loss
$\protect\gamma $. (a) $J=1$, (b) $J=\sqrt{2}$, (c) $J=\sqrt{1+\sqrt{2}}\approx 1.554$, (d) $J=2$. Other parameters are $\protect\kappa =1$, $\Phi =\pi/2$.}
\label{fig4}
\end{figure}

For $J^{2}>J_{\mathrm{c,1}}^{2}$, the $\mathcal{PT}$ phase transition occurs
only once as $\gamma $ increases. $\gamma <\gamma _{\mathrm{c,2,-}}$ is the
exact $\mathcal{PT}$-symmetric phase (phase I in Fig.~\ref{fig2}); $\gamma _{%
\mathrm{c,2,-}}<\gamma <\gamma _{\mathrm{c,1}}$ is the completely broken $%
\mathcal{PT}$-symmetric phase with two conjugate pairs (phase III in Fig.~%
\ref{fig2}): the energy levels are complex when $\gamma _{\mathrm{c,2,-}%
}<\gamma <\gamma _{\mathrm{c,2,+}}$; the energy levels are purely imaginary
when $\gamma _{\mathrm{c,2,+}}<\gamma <\gamma _{\mathrm{c,1}}$; $\gamma
>\gamma _{\mathrm{c,1}}$ is the partially broken $\mathcal{PT}$-symmetric
phase with one conjugate pair (phase II in Fig.~\ref{fig2}). The energy
levels at large degrees of non-Hermiticity $\gamma \gg J,\kappa $ are in the
broken $\mathcal{PT}$-symmetric region with one pair of real energy levels $%
E\approx \pm \kappa $ and one pair of conjugate pairs $\pm i\sqrt{\gamma
^{2}-\kappa ^{2}-2J^{2}}$ [Fig.~\ref{fig3}(g,h)].

The energy levels for magnetic flux $\Phi =\pi /2$ are depicted in Fig.~\ref%
{fig4}; the structures for different couplings are similar as illustrated in
Fig.~\ref{fig3}(c-g). EP2 $\gamma _{\mathrm{c,1}}$ increases as the coupling
$J$ increases. When $J=\sqrt{2}$, $\gamma _{\mathrm{c,2}}$ appears, $\gamma
_{\mathrm{c,2,-}}=\gamma _{\mathrm{c,2,+}}$. As the coupling $J$ continues
to increase, $\gamma _{\mathrm{c,2,-}}$ and $\gamma _{\mathrm{c,2,+}}$
split, $\gamma _{\mathrm{c,2,-}}$ slightly decreases and $\gamma _{\mathrm{%
c,2,+}}$ increases. When $J=J_{\mathrm{c,1}}=\sqrt{1+\sqrt{2}}$, EP4 occurs
and $\gamma _{\mathrm{c,2,+}}=\gamma _{\mathrm{c,1}}$. For even larger $J$,
the structure of the energy levels remains unchanged. Notably, each EP is a
bifurcation point. A typical bifurcation diagram of the spectrum is
illustrated in Fig.~\ref{fig4}(d). When $\gamma <\gamma _{\mathrm{c,2,-}}$,
the system is in the exact $\mathcal{PT}$-symmetric phase in phase I; in the
region $\gamma _{\mathrm{c,2,-}}<\gamma <\gamma _{\mathrm{c,2,+}}$, the
system has two conjugation pairs in phase III; in the region $\gamma _{%
\mathrm{c,2,+}}<\gamma <\gamma _{\mathrm{c,1}}$, the energy levels are
purely imaginary; in the region $\gamma >\gamma _{\mathrm{c,1}}$, the system
has one pair of real energy and one pair of purely imaginary energy levels
in phase II.

\section{Topology of exceptional points}

The phase rigidity for each eigenstate $\psi $ is a useful measure of the
mixing of different states~\cite{Rotter15}, which is defined by%
\begin{equation}
r=\langle \psi ^{\ast }\left\vert \psi \right\rangle /\langle \psi
\left\vert \psi \right\rangle .
\end{equation}%
In Fig.~\ref{fig5}, we depict the phase rigidity and the scaling law for the four eigenstates of Hamiltonian $H$ [Eq.~(\ref{H})] near the EPs.

In a Hermitian system, the phase rigidity is unity for the real value
eigenstate; in a non-Hermitian system, it varies between unity and zero at
completely separate resonance and at level coalescence. The situation
changes in the presence of magnetic flux; the phase rigidity deviates from
unity because the eigenstates are complex even in a Hermitian system. The
scaling law of phase rigidity for different EPs is distinct, $r\propto
|\gamma -\gamma _{\mathrm{c}}|^{\nu }$, where the exponent $\nu $ reflects
the topology of those EPs. The phase rigidity may not vanish at EPs in the
presence of magnetic flux, where the scaling law transforms to $\left\vert
r-r_{\mathrm{c}}\right\vert \propto |\gamma -\gamma _{\mathrm{c}}|^{\nu }$.
For trivial magnetic flux $\Phi =2m\pi $ ($m\in
\mathbb{Z}
$), the phase rigidity decreases to zero at the EPs.

\begin{figure}[tb]
\includegraphics[ bb=0 0 540 750, width=8.8 cm, clip]{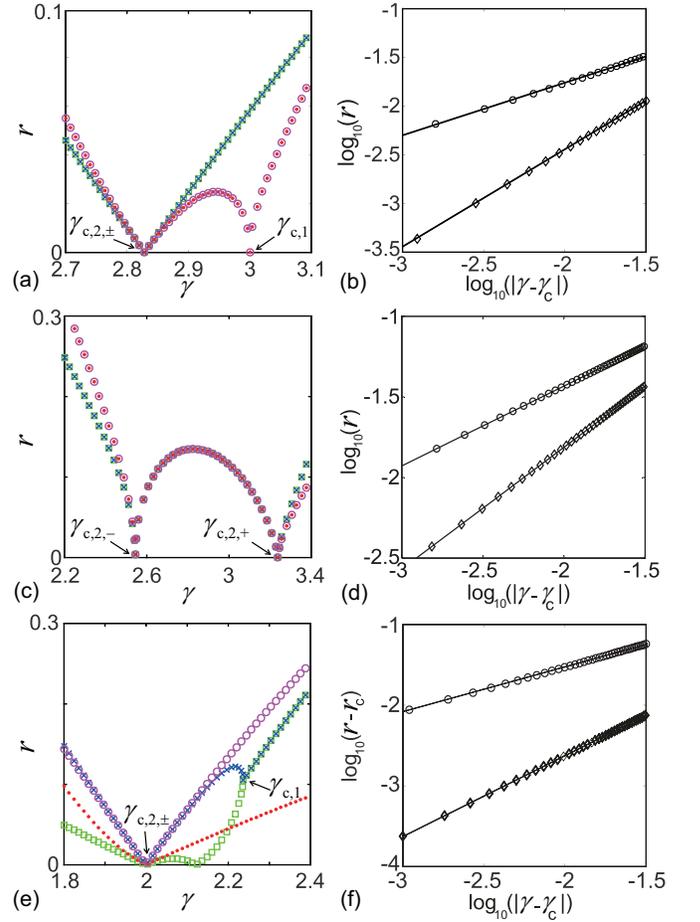}
\caption{(a,c,e) Phase rigidity and (b,d,f) the scaling law when EPs coincide for $\kappa=1$.
(a,b) $J=J_{\mathrm{c,2}}=2$, $\Phi=0$, (c,d) $J=J_{\mathrm{c,1}}=\sqrt{2+\sqrt{5}}$, $\Phi=0$,
(e,f) $J=J_{\mathrm{c,2}}=\sqrt{2}$, $\Phi=\pi/2$; corresponding spectra are in Figs.~\ref{fig3} and~\ref{fig4}. The
fitting exponent approaches 1.0 (0.5) for diamonds (circles) for $\gamma_{\rm{c,2,\pm}}$ ($\gamma_{\rm{c,1}}$) in (b,f); and approaches 0.75 (0.5) for the diamonds (circles) for $\gamma_{\rm{c,2,+}}$ ($\gamma_{\rm{c,2,-}}$) in
(d). Energy levels that coalesced at EPs exhibit identical scaling law in (b,d). (f) is for the green square energy level in (e), the other levels that coalesced at EPs possess identical scaling exponent.} \label%
{fig5}
\end{figure}

The magnetic flux and system spectrum are gauge invariant, but the
eigenstates depend on the local transformation and the phase rigidities
change as local gauge. The scaling exponents are independent of the magnetic
flux, therefore, the phase rigidity can still be used to characterize the
topological properties of the EPs. For the symmetrically coupled asymmetric
dimers of Eq.~(\ref{H}), the separated $\gamma _{\mathrm{c,2,-}}$, $\gamma _{%
\mathrm{c,2,+}}$, and $\gamma _{\mathrm{c,1}}$ are all EP2s. A pair of
two-state coalescence points occurs at $\gamma _{\mathrm{c,2,\pm }}$. The
phase rigidity vanishes at $\gamma _{\mathrm{c,2,-}}$ and $\gamma _{\mathrm{%
c,2,+}} $; the system has two defective eigenstates. Two-state coalescence
occurs at $\gamma _{\mathrm{c,1}}$, where the system has one defective
eigenstate. The phase rigidity at $\gamma _{\mathrm{c,1}}$ does not always
vanishes, rather, it varies as the magnetic flux. The scaling exponents of
separated $\gamma _{\mathrm{c,2,-}}$, $\gamma _{\mathrm{c,2,+}}$, and $%
\gamma _{\mathrm{c,1}}$ are identical; they are all equal to $1/2$ because
of the square root relation of EP2. The fractional exponents indicate the
eigenstates switch and the nontrivial geometric phases are accumulate when
dynamically encircling the EPs~\cite{WDHeissEP,Doppler,Xu}. Encircling each
separated $\gamma _{\mathrm{c,2,-}}$, $\gamma _{\mathrm{c,2,+}}$, and $%
\gamma _{\mathrm{c,1}}$ with two circles, the accumulated geometric phase is
$\pm \pi $. Therefore, the eigenstates return to their initial values after
encircling EP2 with four circles. The accumulated geometric phases are
independent of the magnetic flux.

The topology of EPs changes when they coincide. When the coupling is $J_{%
\mathrm{c,2}}$, $\gamma _{\mathrm{c,2,-}}$ and $\gamma _{\mathrm{c,2,+}}$
coincide ($\gamma _{\mathrm{c,2,-}}=\gamma _{\mathrm{c,2,+}}$) as shown in
Fig.~\ref{fig3}(d). The accumulated geometric phases are related to the
magnetic flux; the summation of geometric phases is invariant, being $\pm
2\pi $ with a pair of symmetric levels ($E_{+,\pm }+E_{-,\pm }=0$) when
encircling $\gamma _{\mathrm{c,2,\pm }}$ for one circle. The scaling
exponent at $\gamma _{\mathrm{c,2,\pm }}$ is $1$. The accumulated geometric
phase is $\pm \pi $ when encircling $\gamma _{\mathrm{c,1}}$ with two
circles; the scaling exponent at $\gamma _{\mathrm{c,1}}$ is $1/2$. When the
coupling is $J_{\mathrm{c,1}}$, $\gamma _{\mathrm{c,2,+}}$ and $\gamma _{%
\mathrm{c,1}}$ coincide ($\gamma _{\mathrm{c,2,+}}=\gamma _{\mathrm{c,1}}$),
as depicted in Fig.~\ref{fig3}(f). All four energy levels coalesce at energy
zero and a high-order EP4 appears, which has an exponent of $3/4$. The
exponent indicates that four circles are needed to make the energy levels
return to their initial values when circling the EP4 in the parameter space.
A geometric phase of $\pm 3\pi $ is accumulated for the eigenstates after
encircling the EP4 with four circles. Therefore, the eigenstates return to
their initial values after encircling the EP4 with eight circles. At EP2 $%
\gamma _{\mathrm{c,2,-}}$, a geometric phase of $\pm \pi $ is accumulated
when encircling $\gamma _{\mathrm{c,2,-}}$ with two circles; the scaling
exponent is $1/2$. In Fig.~\ref{fig5}, we depict the phase rigidity and the
scaling law near the EPs when they coincide. At coupling $J_{\mathrm{c,1}}$,
the phase rigidities are all zero at EPs and the geometric phases are
independent of magnetic flux.

\begin{figure}[tb]
\includegraphics[ bb=0 0 520 500, width=8.6 cm, clip]{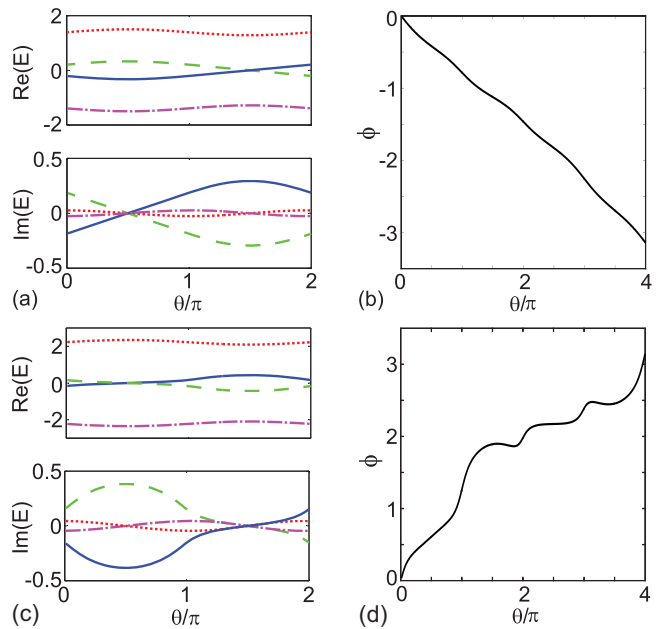}
\caption{Encircling $\protect\gamma_{\rm{c,1}}$ at the
$\protect\delta$-$\protect\kappa$ plane, $\delta =\rho \cos \theta $ and $\kappa =1+\rho \sin \theta $. (a,c) Energy levels, (b,d) accumulated geometric phase. The upper panel is for $J=0.5$, $\gamma=3/4$; the lower panel
is for $J=1.5$, $\gamma=5/4$. Other parameters are $\Phi =0$, $\protect\rho =0.1$.}
\label{fig6}
\end{figure}

\begin{figure}[tb]
\includegraphics[ bb=0 0 370 540, width=8.6 cm, clip]{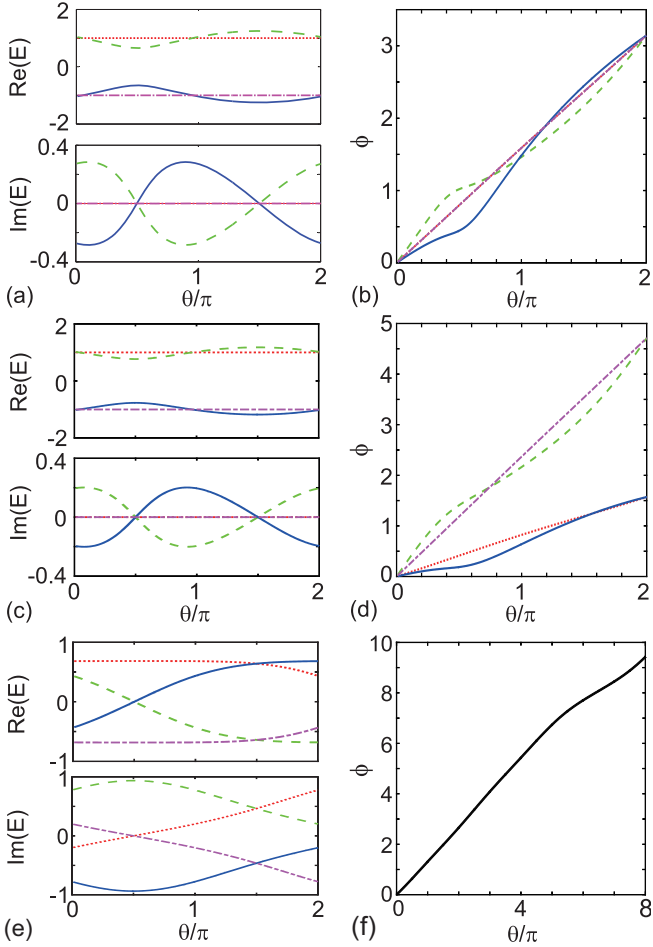}
\caption{Encircling $\protect\gamma_{\rm{c,2,+}}$ at the
$\protect\delta$-$\protect\gamma$ plane, $\delta =\rho \cos \theta $ and $\gamma =\gamma _{\mathrm{c}}+\rho \sin \theta $. (a,c,e) Energy levels, (b,d,f) accumulated
geometric phase. The upper panel is for $\Phi =0$, $J=2$, $\gamma_{\mathrm{c}}=2\sqrt{2}$; the middle panel is for $\Phi =\protect\pi /2$, $J=\protect\sqrt{2}$, $\gamma_{\mathrm{c}}=2$; the lower panel is for $J=1$, $\Phi =\protect\pi $, $\gamma_{\mathrm{c}}=2$. Other parameters are $\kappa=1$, $\protect\rho =0.1$.}
\label{fig7}
\end{figure}

When the coupling $J$ is zero, the system has one critical EP2 at $\gamma
=\kappa $. When the coupling $J$ is switched on, the central two levels
coalesce at EP2 $\gamma _{\mathrm{c,1}}$, which decreases as coupling $J$
increases to $J=\kappa $ and increases for $J>\kappa $ (Fig.~\ref{fig3}).
These are observed in Fig.~\ref{fig3}(a-c). $J=\kappa $, $\gamma =0$ is a
diabolic point; the chirality switches when the system parameter crosses
this point. The chiriality indicates the flow direction, the flow starts
from the gain to the loss in clockwise direction or in counterclockwise
direcion. On the two sides of $J=\kappa $, the coalescence state has two
difference chiralities. In Fig.~\ref{fig3}(a), the coalescence state is $%
(1,i,-1/2,-i/2)^{T}$, which has right chirality; in Fig.~\ref{fig3}(c), the
coalescence state is $(1,-i,-3/2,3i/2)^{T}$, which has left chirality. The
chirality difference of the EPs is reflected from the geometric phase when
encircling the EP2 $\gamma _{\mathrm{c,1}}$, as depicted in Fig.~\ref{fig6}.

We consider a detuning $\delta $ on the gain and loss resonators, where the
Hamiltonian of Eq.~\ref{H} is modified to $H_{1,1}=-\left( \delta +i\gamma
\right) $ and $H_{2,2}=\left( \delta +i\gamma \right) $. The system
parameters encircle the EP in the $\delta $-$\kappa $ parameter plane
according to $\delta =\rho \cos \theta $ and $\kappa =1+\rho \sin \theta $,
where $\rho $ is the radius and $\theta $ is the argument of the encircling.
In the plots, we notice that the central two levels switch after $\theta $
varies $2\pi $; thus, $\theta $ must be $4\pi $ to make the energy levels
return to their initial values. The Berry connection $i\langle \varphi
_{j}\left\vert \partial \right\vert \psi _{j}\rangle /\partial k$ is
complex, and is defined by the eigenstate $\psi _{j}$ of $H$ with eigenvalue
$\varepsilon _{j}$ and corresponding eigenstate $\varphi _{j}$ of $%
H^{\dagger }$ with eigenvalue $\varepsilon _{j}^{\ast }$~\cite%
{CTChanPRB,FengSR}. The complex Berry phase is $(\oint \langle \varphi
_{j}\left\vert \frac{i\mathrm{d}}{\mathrm{d}k}\right\vert \psi _{j}\rangle
\mathrm{d}k)$ for energy level $j$ after the parameter has slowly swept
entire circles in the parameter space. The geometric phases acquired after
encircling $\gamma _{\mathrm{c,1}}$ with two circles in the counterclockwise
direction (varying $\theta $ from $0$ to $4\pi $) are $-\pi $ for both two
central levels in Fig.~\ref{fig6}(b) of the situation shown in Fig.~\ref%
{fig3}(a), where EP2 $\gamma _{\mathrm{c,1}}$ has right chirality;
correspondingly, for the situation in Fig.~\ref{fig3}(c), the geometric
phases acquired after encircling $\gamma _{\mathrm{c,1}}$ with two circles
are $\pi $ as depicted in Fig.~\ref{fig6}(d), where EP2 $\gamma _{\mathrm{c,1%
}}$ has left chirality. Figure~\ref{fig6}(b,d) reflect the chirality
difference of $\gamma _{\mathrm{c,1}}$. For the two central eigenstates,
four circles are required to return the eigenstates to their initial values,
the top and bottom levels return to themselves when the system parameter
encircles $\gamma _{\mathrm{c,1}}$ with one circle.

In the strong coupling region $J\geqslant 2\kappa $, the system experiences
more than one EP as $\gamma $ increases from zero. The first EP is an EP2 $%
\gamma _{\mathrm{c,2,-}}$ with a pair of two-state coalescences: the upper
and lower two levels coalesce. At $J=2\kappa $, the system does not break $%
\mathcal{PT}$ symmetry when system parameter crosses the first EP ($\gamma _{%
\mathrm{c,2,-}}=\gamma _{\mathrm{c,2,+}}$). At regions $\gamma <\gamma _{%
\mathrm{c,2,\pm }}$ or $\gamma >\gamma _{\mathrm{c,2,\pm }}$, the
eigenvalues are all real. The system at the EP2 $\gamma _{\mathrm{c,2,\pm }}$
has a pair of two-state coalescences at $-1$ and $+1$, respectively. To
reveal the topology of $\gamma _{\mathrm{c,2,\pm }}$, we consider $\delta
=\rho \cos \theta $ and $\gamma =\gamma _{\mathrm{c,2,\pm }}+\rho \sin
\theta $ in the $\delta $-$\gamma $ parameter plane. Figure~\ref{fig7}(a-d)
depict the energy levels, the connections, and the accumulated geometric
phases when encircling $\gamma _{\mathrm{c,2,\pm }}$ in the $\delta $-$%
\gamma $ parameter plane for the situations of $\Phi =0$ and $\pi /2$,
respectively. From the energy levels in Fig.~\ref{fig7}(a,c), we notice that
two levels are pinned to $\pm 1$; and the other two levels return to
themselves after $\theta $ varies $2\pi $. Therefore, all eigenstates return
to their initial values after the system parameter encircles the EP with one
round. The real parts of the Berry phases are depicted in Fig.~\ref{fig7}%
(b,d). We notice that the accumulated geometric phases are all $\pi $ for $%
\Phi =0$. The nontrivial magnetic flux in the system changes the Berry
connection and the Berry phase. In the $\Phi =\pi /2$ case, the levels with
positive energy accumulate phases of $\pi /2$; the levels with negative
energy accumulate phases of $3\pi /2$; a pair of opposite energy levels
accumulate a total phase of $2\pi $. The EP4 is at $\Phi =\pi $, $\kappa
=J=1 $, where all four energy levels coalesce at energy zero. From Fig.~\ref%
{fig7}(e), we notice that all four levels return to their initial values
after encircling the EP4 with four complete circles; the geometric phase
acquired is $3\pi $ as shown in Fig.~\ref{fig7}(f). Thus, to make the
eigenstate recover its initial values, another four circles must encircle
EP4.

\section{Conclusion}

We investigate a $\mathcal{PT}$-symmetric system of symmetrically coupled
asymmetric dimers. The nonreciprocal coupling induces effective magnetic
flux, which provides an extra degree of freedom, which helps with control of
the $\mathcal{PT}$ phase transition. The effective magnetic flux is
favorable for the realization of high-order EPs. The EPs and their topology
are far richer than those of a $\mathcal{PT}$-symmetric dimer. Two types of
EP2s exist, which possess one and two defective eigenstates, respectively.
When EPs coincide, the topology of EPs changes; this is reflected by the
scaling exponents and the geometric phases. Although the magnetic flux
changes the eigenstates and the phase rigidity, the phase rigidities are
always zero for the EP2s with a pair of two-state coalescences. The scaling
exponent is independent of the magnetic flux but the geometric phase varies
as magnetic flux when EPs has scaling exponent $1$. The EP4 is a triple
point of different phases; it appears when EP2s with one or two defective
eigenstates coincide. The perturbation around EP4 can result in a
fourth-root mode frequency splitting; the response of mode splitting is
highly sensitive to the cavity frequency perturbation.

\section*{Acknowledgements}

We acknowledge the support of National Natural Science Foundation of China
(Grant No. 11605094) and the Tianjin Natural Science Foundation (Grant No.
16JCYBJC40800).


\begin{thebibliography}{99}
\bibitem{Bender} C. M. Bender and S. Boettcher, Phys. Rev. Lett. \textbf{80}%
, 5243 (1998).

\bibitem{Dorey} P. Dorey, C. Dunning, and R. Tateo, J. Phys. A \textbf{34},
5679 (2001).

\bibitem{Bender02} C. M. Bender, D. C. Brody, and H. F. Jones, Phys. Rev.
Lett. \textbf{89}, 270401 (2002).

\bibitem{Ali} A. Mostafazadeh, J. Math. Phys. \textbf{43}, 205 (2002).

\bibitem{Jones} H. F. Jones, J. Phys. A \textbf{38}, 1741 (2005).

\bibitem{Znojil} M. Znojil, Phys. Rev. D \textbf{78}, 025026 (2008).

\bibitem{LJin09} L. Jin and Z. Song, Phys. Rev. A \textbf{80}, 052107 (2009).

\bibitem{LJin10} L. Jin and Z. Song, Phys. Rev. A \textbf{81}, 032109 (2010).

\bibitem{Witthaut} D. Witthaut, F. Trimborn, H. Hennig, G. Kordas, T.
Geisel, and S. Wimberger, Phys. Rev. A \textbf{83}, 063608 (2011).

\bibitem{YNJ} Y. N. Joglekar and J. L. Barnett, Phys. Rev. A \textbf{84},
024103 (2011).

\bibitem{AR} A. Ruschhaupt, F. Delgado, and J. G. Muga, J. Phys. A: Math.
Gen. \textbf{38}, L171 (2005).

\bibitem{OL} R. El-Ganainy, K. G. Makris, D. N. Christodoulides, and Z. H.
Musslimani, Opt. Lett. \textbf{32}, 2632 (2007).

\bibitem{Musslimani} Z. H. Musslimani, K. G. Makris, R. El-Ganainy, and D.
N. Christodoulides, Phys. Rev. Lett. \textbf{100}, 030402 (2008).

\bibitem{Klaiman} S. Klaiman, U. G\"{u}nther, and N. Moiseyev, Phys. Rev.
Lett. \textbf{101}, 080402 (2008).

\bibitem{Rotter09} I. Rotter, J. Phys. A: Math. Theor. \textbf{42}, 153001
(2009).

\bibitem{Rotter15} I. Rotter and J. P. Bird, Rep. Prog. Phys. \textbf{78},
114001 (2015).

\bibitem{AGuo} A. Guo, G. J. Salamo, D. Duchesne, R. Morandotti, M.
Volatier-Ravat, V. Aimez, G. A. Siviloglou, and D. N. Christodoulides, Phys.
Rev. Lett. \textbf{103}, 093902 (2009).

\bibitem{CERuter} C. E. R\"{u}ter, K. G. Makris, R. El-Ganainy, D. N.
Christodoulides, M. Segev, and D. Kip, Nat. Phys. \textbf{6}, 192 (2010).

\bibitem{Jing} H. Jing, S. K. \"{O}zdemir, X.-Y. L\"{u}, J. Zhang, L. Yang,
and F. Nori, Phys. Rev. Lett. 113, 053604 (2014).

\bibitem{PengNP} B. Peng, S. K. \"{O}zdemir, F. Lei, F. Monifi, M.
Gianfreda, G. L. Long, S. Fan, F. Nori, C. M. Bender and L. Yang, Nat. Phys.
\textbf{10}, 394 (2014).

\bibitem{PengScience} B. Peng, S. K. \"{O}zdemir, S. Rotter, H. Yilmaz, M.
Liertzer, F. Monifi, C. M. Bender, F. Nori, and L. Yang, Science, \textbf{346%
}, 328 (2014).


\bibitem{FengScience} L. Feng, Z. J. Wong, R.-M. Ma, Y. Wang, X. Zhang, Science \textbf{346}, 972 (2014).

\bibitem{Chang} L. Chang, X. Jiang, S. Hua, C. Yang, J.Wen, L. Jiang, G. Li,
G. Wang, and M. Xiao, Nat. Photon. \textbf{8}, 524 (2014).

\bibitem{Feng} L. Feng, Y.-L. Xu, W. S. Fegadolli, M.-H. Lu, J. E. B.
Oliveira, V. R. Almeida, Y.-F. Chen, and A. Scherer. Nat. Mater. 12, \textbf{%
108} (2013).

\bibitem{GraefePRA} E.-M. Graefe and H. F. Jones, Phys. Rev. A \textbf{84},
013818 (2011).

\bibitem{YNJAA} C. H. Liang, D. D. Scott, and Y. N. Joglekar, Phys. Rev. A
\textbf{89}, 030102(R) (2014).

\bibitem{Wiersig} J. Wiersig, Phys. Rev. Lett. \textbf{112}, 203901 (2014).

\bibitem{ChenSensing} W. Chen, S. K. \"{O}zdemir, G. Zhao, J. Wiersig, and
L. Yang, Nature \textbf{548}, 192 (2017).

\bibitem{Dembowski} C. Dembowski, B. Dietz, H.-D. Gr\"{a}f, H. L. Harney, A.
Heine, W. D. Heiss, and A. Richter, Phys. Rev. E 69, 056216 (2004).

\bibitem{Uzdin} R. Uzdin, A. Mailybaev, and N. Moiseyev, J. Phys. A \textbf{44}, 435302 (2011).

\bibitem{WDHeissEP} W. D. Heiss, J. Phys. A: Math. Theor. \textbf{45},
444016 (2012).


\bibitem{Menke} H. Menke, M. Klett, H. Cartarius, J. Main, and G. Wunner,
Phys. Rev. A \textbf{93}, 013401 (2016).


\bibitem{Doppler} J. Doppler, A. A. Mailybaev, J. B\"{o}hm, U. Kuhl, A.
Girschik, F. Libisch, T. J. Milburn, P. Rabl, N. Moiseyev, and S. Rotter,
Nature \textbf{537}, 76 (2016).

\bibitem{Xu} H. Xu, D. Mason, L, Jiang, and J. G. E. Harris, Nature \textbf{%
537}, 80 (2016).

\bibitem{CTChenPRX} K. Ding, G. Ma, M. Xiao, Z. Q. Zhang, and C. T. Chan,
Phys. Rev. X \textbf{6}, 021007 (2016).

\bibitem{HodaeiSensing} H. Hodaei, A. U. Hassan, S. Wittek, H.
Garcia-Gracia1, R. El-Ganainy, D. N. Christodoulides, and M. Khajavikhan
Nature \textbf{548}, 187 (2017).

\bibitem{Fang} K. Fang, Z. Yu and S. Fan, Phys. Rev. Lett. \textbf{108},
153901 (2012).

\bibitem{Fang2} K. Fang, Z. Yu and S. Fan, Nat. Photon. \textbf{6}, 782
(2012).

\bibitem{Tzuang} L. D. Tzuang, K. Fang, P. Nussenzveig, S. Fan, and M.
Lipson, Nat. Photon. \textbf{8}, 701 (2014).

\bibitem{Hafezi} M. Hafezi and P. Rabl, Opt. Exp. \textbf{20}, 7672 (2012).

\bibitem{Li} E. Li, B. J. Eggleton, K. Fang, and S. Fan, Nat. Commun.
\textbf{5}, 3225 (2013).


\bibitem{MHafezi} M. Hafezi, Int. J. Mod. Phys. B \textbf{28}, 1441002
(2014).

\bibitem{HafeziPRL} M. Hafezi, Phys. Rev. Lett. \textbf{112}, 210405 (2014).

\bibitem{Lee} T. E. Lee, Phys. Rev. Lett. \textbf{116}, 133903 (2016).

\bibitem{CTChanPRB} K. Ding, Z. Q. Zhang, and C. T. Chan, Phys. Rev. B
\textbf{92}, 235310 (2015).

\bibitem{FengSR} H. Zhao, S. Longhi, and L. Feng, Sci. Rep. \textbf{5},
17022 (2015).
\end{thebibliography}
\end{document}